\documentclass[12pt]{iopart}

	\usepackage[utf8]{inputenc}  

	\usepackage[T1]{fontenc}  

\expandafter\let\csname equation*\endcsname\relax
  \expandafter\let\csname endequation*\endcsname\relax

	\usepackage[intlimits]{amsmath}		

	\usepackage{amssymb}

	\usepackage{bbold}

	\usepackage{dsfont}      

	\usepackage{fixmath}     

	\usepackage[caption=false]{subfig}

	\usepackage{graphicx}

	\usepackage{ifthen}
	
	\usepackage{epstopdf}

	\usepackage{braket}   
	
	\usepackage[babel,english=american]{csquotes} 
	
	\usepackage{gnuplottex}        

	\usepackage[final,colorlinks=true,linkcolor=blue,citecolor=blue]{hyperref}    

  \newcommand{\ii}{\mathrm{i}}

  \newcommand{\abs}[1]{\left\lvert{#1}\right\rvert}
  
  \newcommand*{\defeq}{\mathrel{\vcenter{\baselineskip0.5ex \lineskiplimit0pt
                     \hbox{\scriptsize.}\hbox{\scriptsize.}}}%
                     =}

  \newlength{\bracewidth}

\newcommand*\density[1]{\hat{\rho}_{#1}}

	\newcommand*\Einchar{E}
	\newcommand*\Einplus[2]{\hat{\Einchar}^{(+)}_{#1}\ifthenelse{\equal{\unexpanded{#2}}{}}{}{(\tout{#2})}}
	\newcommand*\Einminus[2]{\hat{\Einchar}^{(-)}_{#1}\ifthenelse{\equal{\unexpanded{#2}}{}}{}{(\tout{#2})}}

\newcommand\tout[1]{t_{#1}}

\newcommand{\aref}[1]{\hyperref[1]{Appendix~\ref{#1}}}
\newcommand{\difd}{\mathrm{d}}

\newcommand{\avg}[1]{\left\langle #1 \right\rangle}

\renewcommand{\abs}[1]{\left| #1 \right|}
\newcommand{\abssq}[1]{\left| #1 \right|^2}

\newcommand{\idone}{\mathbb{1}}

\newcommand{\Ccurly}{\mathcal{C}}
\newcommand{\Tcurly}{\mathcal{T}}

\newcommand{\Umcp}{\mathcal{U}_{\text{prep}}}
\newcommand{\Pmc}{\mathcal{P}}
\newcommand{\Mmc}{\mathcal{M}}

\DeclareMathOperator{\diag}{diag}
\DeclareMathOperator{\adiag}{antidiag}

\begin{document}
\title{Multipath Correlation Interference and Controlled-NOT Gate Simulation \\with a Thermal Source}
\author{Vincenzo Tamma$^{1,2}$ and Johannes Seiler$^1$}

\address{$^1$Institut f\"{u}r Quantenphysik and Center for Integrated Quantum Science and Technology (IQ\textsuperscript{ST}), Universität Ulm, D-89069 Ulm, Germany}

\address{$^2$Author to whom any correspondence should be addressed.}

\ead{vincenzo.tamma@uni-ulm.de}

\begin{abstract}
We theoretically demonstrate a counter-intuitive phenomenon in optical interferometry with a thermal source: the emergence of second-order interference between two pairs of correlated optical paths even if the time delay imprinted by each path in one pair with respect to each path in the other pair is much larger than the source coherence time. This fundamental effect could be useful for experimental simulations of small-scale quantum circuits and of $100\%$-visibility correlations typical of entangled states of a large number of qubits, with possible applications in high-precision metrology and imaging. As an example, we demonstrate the polarization-encoded simulation of the operation of the quantum logic gate known as controlled-NOT gate. 
\end{abstract}
\submitto{\NJP}
\maketitle


\section{Motivation}
The Hanbury Brown and Twiss (HBT) effect \cite{HBT, HBT2} discovered in $1956$ was at the heart of the development of the field of quantum optics. 
Indeed, this discovery led to numerous remarkable multiphoton experiments which have not only deepened our fundamental understanding of multiphoton interference \cite{Glauber1963, Glauber2006, Shih1986, HOM, Franson1989, tammalaibacher2014, tammalaibacherthermal2014} but have also led to numerous applications in information processing \cite{nielsen, pan, MBCS, tamma2014, JOMtammalaibacher, 
tamma_multi-boson_2015} and imaging \cite{ShihBook2011, pittmanshih, Bennink, Valencia, Ferri, Oppel_2012, Pearce2015}.


The HBT effect fundamentally reveals the second-order coherence of a thermal source. For example, second-order temporal correlations can be measured  after the interaction of multi-mode thermal radiation of given bandwidth $\Delta\omega$ with a beam splitter: two detectors at the beam splitter output ports have twice the chance to be triggered at  equal times than with a relative time delay longer than the coherence time $1/\Delta\omega$ of the thermal field. 

It is interesting to modify the described HBT scheme by adding two unbalanced Mach-Zehnder interferometers at the beam splitter output ports, the ``control'' port $\Ccurly$ and the ``target'' $\Tcurly$, as depicted in \fref{fig:2 mode unitary transformation polarization correlation example}(a). We further consider path lengths $L \defeq L_C \cong L_T$ and $S \defeq S_C \cong S_T$ such that the time $\mid L- S \mid/c$, with $c$ the speed of light, is much larger than the coherence time of the source. Can we observe second order interference by performing correlation measurements at equal detection times at the output of the two Mach-Zehnder interferometers?

Interestingly, in this paper we show that second-order interference between the two pairs $(L_C, L_T)$ and $(S_C, S_T)$ of optical paths (\textit{multipath correlation interference}) can be observed even if the time delay $\mid L- S \mid/c$ imprinted by each path in the pair $(L_{C},L_{T})$ with respect to each path in the pair $(S_{C},S_{T})$ is much beyond the source coherence time. Furthermore, we describe the similarities and differences between this second-order interference effect and the one demonstrated by Franson \cite{Franson1989} in $1989$ by using a two-photon entangled source.

 The fundamental interference phenomenon described here is also of interest in view of the recent studies of simulations of quantum logic operations and entanglement correlations using classical light \cite{
PhysRevA.57.R1477, PhysRevA.63.062302, Lee2002, Lee2004, Fu2004, Chen_2011, Kagalwala2013, Peng2015}. In particular, by considering the interferometric scheme in \fref{fig:2 mode unitary transformation polarization correlation example}(b), we demonstrate how this interference effect is able to simulate the result of a  controlled-NOT (CNOT) logic operation \cite{nielsen, lomonaco, pittman, OBrien2003, Gasparoni2004, Okamoto2005, pnasknill, sanaka}. 

The paper is outlined as following: we demonstrate how multipath correlation interference can be observed with a thermal source in section \ref{sec:intdesc}; we apply this novel phenomenon to the simulation of a CNOT gate operation in section \ref{sec:clc}; we show how this second-order interference effect can be generalized to interferometers based on arbitrary-order correlation measurements in section \ref{sec:mpn}; and we conclude with discussions in section \ref{sec:disc}.

\section{Multipath Correlation Interference} \label{sec:intdesc}
\begin{figure}[tb]
\begin{flushleft}
\hspace{3.15cm} (a)
\end{flushleft}
\vspace{-1.2cm}
\begin{center}
\includegraphics[scale=1]{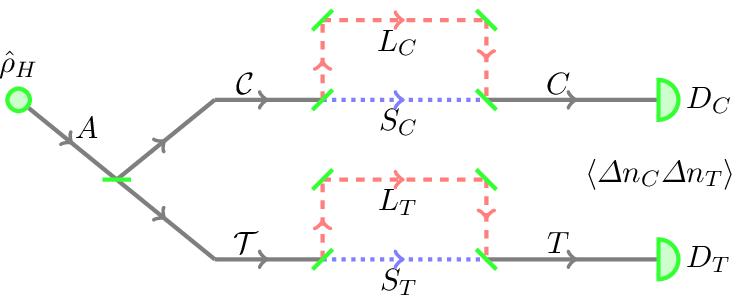}
\begin{flushleft}
\hspace{3.15cm} (b)
\end{flushleft} 
\vspace{-.8cm}
   \includegraphics[scale=1]{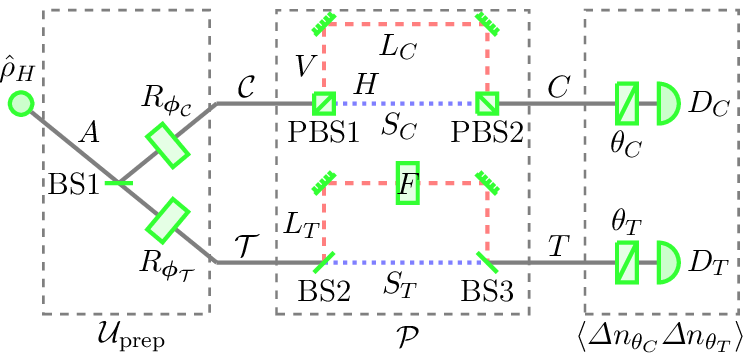}
  \caption[]{(a) Modified HBT interferometer with a thermal source where two unbalanced Mach-Zehnder interferometers are placed at the output channel of the first beam splitter and the correlation between the photon-number fluctuations is measured at the interferometer output. (b) Interferometer simulating a CNOT-gate operation: 1) The first transformation $\Umcp$ prepares the initial polarization state; 2) By using two unbalanced interferometers a polarization-dependent transformation $\Pmc$ is implemented; 3) At the interferometer output the polarization-dependent correlation between the photon-number fluctuations is measured.
  }
  \label{fig:2 mode unitary transformation polarization correlation example}
\end{center}
\end{figure}

We consider here the interferometer in \fref{fig:2 mode unitary transformation polarization correlation example}(a).   
The interferometer has only one source, which generates in the input port $A$ thermal light with a given horizontal polarization $H$. Therefore, the input state is described by  \cite{Glauber2007,Mandel1995}
\begin{align}
  \density{H}
  & =
  \int \left[\prod_{\omega} \difd^2 \alpha_{\omega,H}\right] P_{\hat{\rho}_{H}}(\{\alpha_{\omega,H}\}) \bigotimes_{\omega} \ket{\alpha_{\omega,H}}_{A} \! \bra{\alpha_{\omega,H}} 
\label{eq:Density_op_Thermal_B_Two_Mode},
\end{align}
with the Glauber-Sudarshan probability distribution \cite{Glauber1963, Sudarshan1963}
\begin{equation}
 P_{\hat{\rho}_{H}}(\{\alpha_{\omega,H}\})=\prod_{\omega}\frac{1}{\pi\,\overline{n}_{\omega}}\exp\left(-\frac{\abssq{\alpha_{\omega,H}}}{\overline{n}_{\omega} }\right),
 \nonumber
\end{equation}
where $\overline{n}_{\omega}$ is the average photon number at frequency $\omega$. For simplicity, but without losing generality, we consider a Gaussian frequency distribution \cite{Glauber2007}
\begin{align}
  \overline{n}_{\omega} &= \overline{r} \frac{1}{\sqrt{2\pi}\Delta \omega} \, \exp\left\lbrace -\frac{\left(\omega - \omega_{0} \right)^2}{2 \Delta\omega^{2}} \right\rbrace
  \nonumber
\label{eq:C02_014_Text},
\end{align}
with the mean photon rate $\overline{r}$, average frequency $\omega_{0}$ and spectral width $\Delta \omega$. 

At the interferometer output correlation measurements in the photon-number fluctuations
\begin{equation}
\Delta n_{d} = n_{d} - \avg{n_{d}}
\label{eq:pnf_def}
\end{equation}
at the detection time $t_{d}$ around the mean value $\avg{n_{d}}$, with $d=C,T$, are performed by using either photon-number resolving detectors or single-photon detectors \cite{ShihScully2014, Shih} with integration time $\delta t \ll 1/\Delta \omega$, which are currently available (e.g. $\delta t \sim 10-100 \,ns$).
The expectation value for the product of the  photon-number fluctuations in \eref{eq:pnf_def} at the two output ports is
\cite{ShihScully2014, Shih, Glauber2007}
\begin{align}
\avg{\Delta n_{C}\Delta n_{T}} &= \avg{n_{C}\,n_{T}} - \avg{n_{C}}\avg{n_{T}}\notag \\ 
&\propto G^{(2)}(t_{C},t_{T}) - G^{(1)}(t_{C},t_{C})G^{(1)}(t_{T},t_{T})  = \abssq{G^{(1)}(t_{C},t_{T})}.
\label{eq:corrfluc}
\end{align}
Here we used the properties \cite{Glauber2007}
\begin{equation}
\avg{n_{C}\, n_{T}} \propto G^{(2)}(t_{C},t_{T}), \notag
\end{equation}
with
\begin{equation}
G^{(2)}(t_{C},t_{T}) = G^{(1)}(t_{C},t_{C})G^{(1)}(t_{T},t_{T}) + \abssq{G^{(1)}(t_{C},t_{T})}
\label{eq:G2_expression_in_G1}
\end{equation}
and
\begin{equation}
\avg{n_{d}} \propto G^{(1)}(t_{d},t_{d}), \notag
\end{equation}
where $G^{(1)}$ and $G^{(2)}$ are, respectively, the first and second-order correlation functions \cite{Glauber2007}. Therefore, the outcome \eref{eq:corrfluc} of the correlation measurement does not depend on the ``background'' term $G^{(1)}(t_{C},t_{C})G^{(1)}(t_{T},t_{T})$ in \eref{eq:G2_expression_in_G1}. As we will show, the only relevant term $\abssq{G^{(1)}(t_{C},t_{T})}$ characterizes the second-order interference occurring in the interferometer. Toward this end, we introduce the definition of the first-order correlation function  
\begin{equation}
G^{(1)}(t_{C},t_{T})\defeq\tr \left[\hat{\rho}_{H} {\hat{E}}_{C}^{(-)}\!(t_{C})  {\hat{E}}_{T}^{(+)}\!(t_{T})  \right].
\label{eq:first-order-correlation}
\end{equation}
Here, in the narrow bandwidth approximation,
\begin{equation}
{\hat{E}}_{d}^{(+)}(t_{d})\propto\!\!\int\!\! \difd \omega\; \e^{-i\omega t_{d}}\left[\e^{\ii \omega L_{d}/c} + \e^{\ii \omega S_{d}/c} \right] \;
\hat{a}_{A}^{(H)}(\omega)
\label{eq:atransform3}
\end{equation}
is the electric field operator at the detector $d=C,T$ in terms of the frequency-dependent annihilation operator $\hat{a}_{A}(\omega)$ at the only input port $A$ where a source is placed; while ${\hat{E}}^{(-)}_{d}\!(t_{d})$ is its respective Hermitian conjugate. The factors $\e^{\ii \omega L_{d}/c}$ and $\e^{\ii \omega S_{d}/c}$ describe the propagation through the paths $L_{d}$ and $S_{d}$, respectively. 
By using \eref{eq:first-order-correlation} and \eref{eq:atransform3}, and defining the \textit{effective} detection times  
\begin{equation}
t_d^{(l_d)} \defeq t_{d} - \frac{l_{d}}{c},
\label{eq:def_eff_det_times}
\end{equation}
 with $l_d =L_{d},S_{d}$, we show in \ref{sec:appA} that the expectation value in \eref{eq:corrfluc} can be written as  
\begin{equation}
\avg{\Delta n_{C} \Delta n_{T}}\propto \abssq{G^{(1)}(t_{C},t_{T})}=\abssq{\sum_{l_{C},l_{T}} G^{(l_{C},l_{T})}(t_{C},t_{T})}.
\label{eq:corrfluct}
\end{equation}
Here, each interfering term
\begin{equation}
G^{(l_{C},l_{T})}(t_{C},t_{T}) = \ii \, a\,\overline{r}\, \e^{\ii \omega_{0}[t_{C}^{(l_{C})}-t_{T}^{(l_{T})}]}\e^{-[t_C^{(l_C)}- t_T^{(l_T)}]^2 \Delta \omega^2 /2},
\label{eq:Gll}
\end{equation} 
with $a$ constant, describing the contribution of the corresponding path pair ($l_{C},l_{T}$), 
 depends on the time delay $t_C^{(l_C)}- t_T^{(l_T)}$ between the detected photons in the two output ports before the propagation through the two paths $l_{C}$ and $l_{T}$, respectively. In a standard HBT experiment only a fixed pair of paths can contribute to the measurement. Differently, here all the four possible pairs $(l_{C},l_{T})=(L_{C},L_{T}),(S_{C},S_{T}),(L_{C},S_{T}),(S_{C},L_{T})$ of paths can lead to a joint detection and, in principle, can interfere as in \eref{eq:corrfluct}.

Now, we consider the following conditions for the delays between the effective detection times defined in \eref{eq:def_eff_det_times}:
\begin{align}
\abs{t_{C}^{(L_{C})}-t_{T}^{(L_{T})}} \ll 1/\Delta\omega, \quad\quad
\abs{t_{C}^{(S_{C})}-t_{T}^{(S_{T})}} \ll 1/\Delta\omega, 
\label{eq:8} 
\end{align}
and
\begin{align}
\abs{t_{C}^{(L_{C})}-t_{T}^{(S_{T})}} \gg 1/\Delta\omega,  \quad\quad
\abs{t_{T}^{(L_{T})}-t_{C}^{(S_{C})}} \gg 1/\Delta\omega, \label{eq:8a}
\end{align}
which, by using \eref{eq:8}, corresponds to differences between the path lengths in each Mach-Zehnder interferometer much larger than the coherence length $c/\Delta\omega$ of the source:
\begin{align}
\abs{L_{C}-S_{C}} \gg c/\Delta\omega,  \quad\quad
\abs{L_{T}-S_{T}} \gg c/\Delta\omega \label{eq:8b}.
\end{align}
The conditions \eref{eq:8} can be simply achieved experimentally, for example,
 in the limit  
 \begin{align}
t_{C} \cong t_{T}  \,\, \Leftrightarrow \,\, \abs{t_{C}-t_{T} }&\ll 1/\Delta\omega \label{eq:Cnot conditions 01c}
\end{align}
of approximately equal detection times with respect to the coherence time $1/\Delta \omega$ 
and approximately equal paths 
 \begin{align}
L \defeq L_{C} \cong L_{T}   \,\, \Leftrightarrow \,\, \abs{L_{C}-L_{T}} \ll c/\Delta\omega, \label{eq:Cnot conditions 02a} 
\quad\quad
S \defeq S_{C} \cong S_{T}   \,\, \Leftrightarrow \,\, \abs{S_{C}-S_{T}} \ll c/\Delta\omega 
\end{align} 
with respect to to the coherence length $ c/\Delta\omega $. By using  \eref{eq:Cnot conditions 02a}, the two conditions \eref{eq:8b} reduce to the single condition
\begin{align}
\abs{L -S} \gg c/\Delta\omega \label{eq:Cnot conditions 02aa}.
\end{align}

In the conditions \eref{eq:8a} and \eref{eq:8} and by using \eref{eq:Gll} the expression in \eref{eq:corrfluct} becomes
\begin{align}
\avg{\Delta n_{C} \Delta n_{T}}
&\propto \abssq{G^{(1)}(t_{C},t_{T})} \cong \abssq{G^{(L_{C},L_{T})}(t_{C},t_{T})+G^{(S_{C},S_{T})}(t_{C},t_{T})}
\nonumber \\ 
&= 2a^2\overline{r}^2 \left(1+ \cos \varphi_{L-S}   \right),
\label{eq:pnf coh gen}
\end{align}
with the relative phase
\begin{align}
\varphi_{L-S}\defeq& \omega_{0}\left[(t_{C}^{(L_{C})} - t_{T}^{(L_{T})})- (t_{C}^{(S_{C})} - t_{T}^{(S_{T})})\right]
= \frac{\omega_{0}}{c}\left[(L_{C}-L_{T})-(S_{C}-S_{T} )\right],\label{eq:limit of omega 0}
\end{align} 
where we used \eref{eq:def_eff_det_times} in the second equality. 
The expectation value \eref{eq:pnf coh gen} depends now on the interference between \textit{only} two contributions  $G^{(L_{C},L_{T})}$ and $G^{(S_{C},S_{T})}$ associated with the pairs of optical paths $(L_C,L_T)$ and $(S_C,S_T)$, respectively. The paths $L_C$ and $S_C$ in the interferometer are correlated with the paths $L_T$ and $S_T$, respectively, and \textit{only} the corresponding path pairs $(L_C,L_T)$ and $(S_C,S_T)$ \textit{interfere}. This interference occurs even if the differences $\abs{L_d - S_d}$, with $d = C,T$, between the path lengths in each Mach-Zehnder interferometer are much larger than the coherence length of the source (see \eref{eq:8b}). Indeed, for a thermal source, the interfering contributions $G^{(L_{C},L_{T})}$ and $G^{(S_{C},S_{T})}$ (see \eref{eq:Gll}) do not depend on the relative path lengths in each Mach-Zehnder interferometer  but \textit{only} on the difference between the delays $t_{C}^{(L_{C})} - t_{T}^{(L_{T})}$ and $t_{C}^{(S_{C})} - t_{T}^{(S_{T})}$ between the detected photons in the two output ports before the propagation through the two pairs $(L_C,L_T)$ and $(S_C,S_T)$ of paths, respectively. Since these time delays are very small compared to the coherence time of the source (see \eref{eq:8}) both pairs $(L_C,L_T)$ and $(S_C,S_T)$ contribute to the observed second-order interference.

It is worthwhile to compare the interferometer described here with the famous Franson interferometer \cite{Franson1989}, where the state at the output of the beam splitter in \fref{fig:2 mode unitary transformation polarization correlation example}(a) is substituted by a two-photon entangled state and the coincidence rate for detecting a photon at the same time in both output ports is measured. 
For a full comparison, we first determine the coincidence rate associated with the absorption of a single photon from the field at each of the two output ports of the interferometer  in  \fref{fig:2 mode unitary transformation polarization correlation example}(a) considered here. This corresponds to measure the standard second-order correlation function \cite{Glauber2007} in \eqref{eq:G2_expression_in_G1} at approximately equal detection times $t_{C}\cong t_{T}$ (see \eref{eq:Cnot conditions 01c}) and in the conditions (\ref{eq:Cnot conditions 02a}) and (\ref{eq:Cnot conditions 02aa}) for the interferometric optical paths. In particular, by adding the product  $G^{(1)}(t_C,t_C)G^{(1)}(t_T,t_T) = 4 \,a^2\,\overline{r}^2$ of the intensities at the two output ports to the second-order interference term $\abssq{G^{(1)}(t_{C}, t_{T})}$ found in \eref{eq:pnf coh gen}, we obtain a second-order correlation function  $$G^{(2)}(t_{C}, t_{T})\propto  3+ \cos\varphi_{L-S}    $$ with visibility $1/3$. This is a crucial difference between the interferometer described here and the Franson interferometer, where the second-order correlation function measured at the output at equal detection times  manifests second-order interference with $100\%$ visibility. Instead, in the interferometer in \fref{fig:2 mode unitary transformation polarization correlation example}(a), $100\%$-visibility interference is only achieved by measuring the correlation \eqref{eq:pnf coh gen} between the fluctuations in the number of detected photons, where the ``background'' constant term  $G^{(1)}(t_C,t_C)G^{(1)}(t_T,t_T)$ is effectively ``subtracted'' from the second order correlation function. 
Therefore, the emergence of this interference effect is very different from the physics of two-photon interference based on energy-time entanglement in the Franson interferometer. 
Indeed, in the Franson interferometer the interference between the two  pairs $(L,L)$ and $(S,S)$ of optical paths emerge from the fact that the two input entangled photons are emitted at the same time and the joint emission time is uncertain in the quantum sense. Therefore, $100\%$-visibility second-order interference can be observed even if the first-order coherence length is much less than the difference $L-S$ between the path lengths. 
In the interferometer described here, instead, neither an entangled source is used nor any entanglement process occurs. 
Therefore, coincidence events do not necessarily correspond to the absorption from the field of photons which entered the interferometer at the same time. In principle, the detected photons could have taken any of the four  possible pairs of paths from the source to the two detectors. Nonetheless, the interference term $\abssq{G^{(1)}(t_{C},t_{T})}$ in the second-order correlation function, emerging  from the measurement of the correlation \eqref{eq:pnf coh gen} between the fluctuations in the number of detected photons as an average over all the possible experimental outcomes, contains only the contribution of two  indistinguishable pairs $(L_C,L_T)$ and $(S_C,S_T)$ of correlated paths  \footnote{Interestingly,  there is, in principle, no upper bound to the difference  $\abs{L-S}$ between the path lengths in \eref{eq:Cnot conditions 02aa} which limit this interference effect for an ideal stationary thermal source. This is of course not the case in ``real world'' experiments.}. 
In particular, the interference pattern \eqref{eq:pnf coh gen} depends on the \textit{difference} $$\varphi_{L-S}=\varphi_{C}-\varphi_{T}$$ in \eref{eq:limit of omega 0} between the relative phases $\varphi_{d}=\omega_{0}(L_{d}-S_{d})/c $, with $d=C,T$, in the two Mach-Zehnder interferometers. Differently, in the Franson interferometer the resulting interference pattern depends on the \textit{sum} of the relative phases in the two Mach-Zehnder interferometers.

\section{CNOT Gate Simulation} \label{sec:clc}

The interference phenomenon based on multipath correlations demonstrated in the previous section can be used to reproduce on-demand correlations in different degrees of freedom without the use of entanglement.
Here, we address, for example, the simulation of a CNOT gate operation by encoding these multipath correlations in the polarization degree of freedom.

For this purpose, we consider the interferometer in \fref{fig:2 mode unitary transformation polarization correlation example}(b). 
The source at the input port $A$ is again described by \eref{eq:Density_op_Thermal_B_Two_Mode}.
The $H$-polarized thermal light impinges on the balanced beam splitter BS1 and, by using half-wave plates, is prepared at the ``control'' port $\Ccurly$ and ``target'' port $\Tcurly$  in two general polarizations $\boldsymbol{\phi}_{\Ccurly} = \left( \cos\phi_{\Ccurly}\quad\sin\phi_{\Ccurly}\right)^T$ and $\boldsymbol{\phi}_{\Tcurly} = \left( \cos\phi_{\Tcurly}\quad\sin\phi_{\Tcurly}\right)^T$, respectively. Here, the $H$ and $V$ polarization directions are indicated by the vectors $(1\quad 0)^T$ and $(0\quad 1)^T$, respectively. 
Thereby, by introducing the polarization rotations
\begin{equation}
R_{\boldsymbol\phi_{\Ccurly,\Tcurly}}\defeq
\begin{pmatrix}
\cos\phi_{\Ccurly,\Tcurly} &  \sin\phi_{\Ccurly,\Tcurly}\\
\sin\phi_{\Ccurly,\Tcurly} & -\cos\phi_{\Ccurly,\Tcurly}
\end{pmatrix},
\nonumber
\label{eq:Urot}
\end{equation}
the interferometer transformation associated with the first part of the interferometer connecting the input ports $A,B$ with the ports $\Ccurly,\Tcurly$ is given by \begin{equation}
\Umcp\defeq
\begin{pmatrix}
R_{\phi_{\Ccurly}} & 0 \\
0 & R_{\phi_{\Tcurly}}
\end{pmatrix}
\mathcal{U}_{\text{BS}}
= 
\frac{1}{\sqrt{2}}
\begin{pmatrix}
R_{\phi_{\Ccurly}} & 0 \\
0 & R_{\phi_{\Tcurly}}
\end{pmatrix}
\begin{pmatrix}
\ii \idone & \idone\\
\idone & \ii \idone
\end{pmatrix}
\label{eq:Uprep},
\end{equation}
where we used the expression for the balanced beam-splitter transformation $\mathcal{U}_{\text{BS}}$,
%
with $\idone \defeq
\diag \left(1,1\right)
\nonumber
$.

The second part of the interferometer consists of a ``control'' interferometer connecting the ports $\Ccurly$ and $C$ and a  ``target'' interferometer connecting the ports $\Tcurly$ and $T$. Therefore, the global interferometric  evolution is described by the two polarization-dependent transformations $\Pmc_{C,\Ccurly}$ and $\Pmc_{T,\Tcurly}$ defining the diagonal matrix 
\begin{equation}
\Pmc\defeq
\diag \left(\Pmc_{C,\Ccurly},\Pmc_{T,\Tcurly}\right).
\label{eq:Ucnot}
\end{equation}
In particular, in the control interferometer, the light in the polarization modes $H$ and $V$ at the output of the first polarizing beam splitter PBS1 acquires the time delays $S_{C}/c$ and $L_{C}/c$, respectively, with $c$ the speed of light, before being recombined at the output of the second polarizing beam splitter PBS2. This leads to the control transformation
\begin{equation}
\Pmc_{C, \Ccurly}\defeq
\diag \left(
\e^{\ii \omega S_{C}/c},\e^{\ii \omega L_{C}/c}
\right).
\nonumber
\label{eq:UCtoC}
\end{equation}
 On the other hand, the light in the target interferometer is coherently split into two different paths and recombined by the balanced beam splitters BS2 and BS3, respectively, independently of the polarization. 
The light polarization is unchanged in the path of length  $S_{T}$. Instead, in the path of length $L_{T}$ the polarization modes $H$ and $V$ are flipped ($H \leftrightarrow V$) by the NOT-gate operation 
$ F\defeq
\adiag \left(1,1\right)
\nonumber
$
implemented by a half-wave plate with axes rotated by $\pi/4$ with respect to the $H$ and $V$ axes. Thus the overall target evolution is described by the transformation
\begin{equation}
 \Pmc_{T, \Tcurly}
\defeq \frac{1}{2}\left(
\e^{\ii \omega L_{T}/c}F+  \e^{\ii \omega S_{T}/c}\idone \right).
\nonumber
\label{eq:CT Tin Tout}
\end{equation}
By using \eref{eq:Uprep} and \eref{eq:Ucnot}, we derive the total interferometer matrix
\begin{align}
\Mmc(\omega)
\defeq \Pmc \cdot \Umcp 
=\frac{1}{\sqrt{2}}
\begin{pmatrix}
\ii \, \Pmc_{C,\Ccurly}\, R_{\phi_{\Ccurly}} & \Pmc_{C,\Ccurly}\, R_{\phi_{\Ccurly}} \\
\Pmc_{T,\Tcurly} \, R_{\phi_{\Tcurly}} & \ii \,\Pmc_{T,\Tcurly}\, R_{\phi_{\Tcurly}}
\end{pmatrix} .
\label{eq:4by4transformation}
\end{align}

We finally address the detection process, consisting of measuring the polarization-dependent correlation between the fluctuations in the number of photons $\Delta n_{{\theta}_{C}}(t_C)$ and $\Delta n_{{\theta}_{T}}(t_T)$ detected at the control and target ports $d=C,T$, respectively, with polarization  $\boldsymbol{\theta}_{d}\defeq\left( \cos\theta_{d}\quad\sin\theta_{d}\right)^T$ at time $t_d$.  The expectation value for the product of the  photon-number fluctuations at the two output ports is
\cite{ShihScully2014, Shih, Glauber2007}
\begin{equation}
\avg{\Delta n_{{\theta}_{C}} \Delta n_{{\theta}_{T}}}\propto\abssq{G^{(1)}_{{\theta}_{C},{\theta}_{T}}(t_{C},t_{T})},
\label{eq:corrfluc2}
\end{equation}
where the first-order correlation function  
\begin{equation}
G^{(1)}_{{\theta_{C}},{\theta_{T}}}(t_{C},t_{T})\defeq\tr \left[\hat{\rho}_{H} \left( \boldsymbol\theta_{C} \cdot \boldsymbol{\hat{E}}_{C}^{(-)}\!(t_{C})  \right)\left( \boldsymbol\theta_{T} \cdot \boldsymbol{\hat{E}}_{T}^{(+)}\!(t_{T}) \right) \right]
\notag
\end{equation}
is now, differently from \eqref{eq:first-order-correlation}, polarization dependent. Here, in the narrow bandwidth approximation and in the $(H,V)$-polarization basis, 
\begin{equation}
\boldsymbol{\hat{E}}_{d}^{(+)}(t_{d})\propto\!\!\int\!\! \difd \omega\; \e^{-i\omega t_{d}}\Mmc_{d,A}(\omega) \;
\begin{pmatrix}
\hat{a}_{A}^{(H)}(\omega) \\ \hat{a}_{A}^{(V)}(\omega)
\end{pmatrix},
\notag
\end{equation}
with the elements $\Mmc_{d,A}=\Mmc_{C,A},\Mmc_{T,A}$ in the first column of the total interferometer matrix \eref{eq:4by4transformation},
is the electric field operator at the detector $d=C,T$ in terms of the frequency-dependent annihilation operators $\hat{a}_{A}^{(H)}(\omega)$ and $\hat{a}_{A}^{(V)}(\omega)$ at the only input port $A$ where the source is placed, while $\boldsymbol{\hat{E}}^{(-)}_{d}\!(t_{d})$ is its respective Hermitian conjugate.  
In \ref{sec:appB} we show, that, in the limits \eref{eq:8a} and \eref{eq:8}, \eref{eq:corrfluc2} becomes 
\begin{align}
\avg{\Delta n_{{\theta}_{C}} \Delta n_{{\theta}_{T}}}
&\propto\abssq{G_{\theta_{C} \theta_{T}}^{(L_{C},L_{T})}(t_{C},t_{T})+G_{\theta_{C} \theta_{T}}^{(S_{C},S_{T})}(t_{C},t_{T})}\nonumber \\ 
&\propto \overline{r}^2 \left|\cos\phi_{\Ccurly}\cos\theta_{C} \cos\left(\phi_{\Tcurly}-\theta_{T}\right) + \e^{\ii \varphi_{L-S}} \sin\phi_{\Ccurly}\sin\theta_{C} \sin\left(\phi_{\Tcurly}+\theta_{T}\right)  \right|^{2},
\label{eq:pnf coh gen 2}
\end{align}
with $\varphi_{L-S}$ defined in (\ref{eq:limit of omega 0}).

Here, similarly to a CNOT gate operation, the control path $S_{C}$ associated with the polarization mode $H$ is correlated with the target path $S_{T}$ where the light polarization remains unchanged; instead, the control path $L_{C}$ associated with the polarization mode $V$ is correlated with the target path $L_{T}$ where a NOT-gate operation occurs. Moreover, analogously to the scheme in \fref{fig:2 mode unitary transformation polarization correlation example}(a), the two pairs $(L_{C},L_{T})$ and $(S_{C},S_{T})$ of optical paths interfere.
 
Let us fully compare now the interferometer in \fref{fig:2 mode unitary transformation polarization correlation example}(b) with a \textit{genuine} CNOT entangling operation on the two-qubit input state $\ket{\phi_{\Ccurly}}_{\Ccurly}\ket{\phi_{\Tcurly}}_{\Tcurly}$, where 
\begin{equation}
\ket{\phi_{\Ccurly}}_{\Ccurly}\defeq\cos\phi_{\Ccurly}\ket{H}_{\Ccurly}+\sin\phi_{\Ccurly}\ket{V}_{\Ccurly},
\notag
\end{equation}
and
\begin{equation}
\ket{\phi_{\Tcurly}}_{\Tcurly}\defeq\cos\phi_{\Tcurly}\ket{H}_{\Tcurly}+\sin\phi_{\Tcurly}\ket{V}_{\Tcurly}
\notag
\end{equation}  
are expressed as superpositions of the polarization states $\ket{H}$ and $\ket{V}$, corresponding to single-photon occupations of the H and V modes, respectively. A CNOT-gate operation on this input state leads to the output entangled state 
\begin{align}
\ket{\psi}_{C,T}=\cos\phi_{\Ccurly}\ket{H}_{C}\ket{\phi_{\Tcurly}}_{T}+\sin\phi_{\Ccurly}\ket{V}_{C}\ket{\phi_{\Tcurly}^{(F)}}_{T},
\notag
\end{align}
where 
\begin{equation}
\ket{\phi_{\Tcurly}^{(F)}}_{T}\defeq\sin\phi_{\Tcurly}\ket{H}_{T} + \cos\phi_{\Tcurly}\ket{V}_{T}.
\nonumber 
\end{equation}
Polarization-correlation measurements over the state $\ket{\psi}_{C,T}$ occur with a probability
\begin{align}
P_{\text{CNOT}}\defeq&\abssq{\braket{\theta_{C},\theta_{T}|\psi}_{C,T}}
=&\left|\cos\phi_{\Ccurly}\cos\theta_{C} \cos\left(\phi_{\Tcurly}-\theta_{T}\right) + \sin\phi_{\Ccurly}\sin\theta_{C} \sin\left(\phi_{\Tcurly}+\theta_{T}\right) \right|^{2}. 
\label{eq:PcNOT}
\end{align}
Comparing \eref{eq:PcNOT} with \eref{eq:pnf coh gen 2} in the limit $\varphi_{L-S}\ll 1$, we obtain
the expectation value
\begin{align}
\avg{\Delta n_{{\theta}_{C}} \Delta n_{{\theta}_{T}}}&\propto \overline{r}^2 P_{\text{CNOT}},
\label{eq:PcNOT2}
\end{align}
which takes into account all the possible outcomes for the product of the photon-number fluctuations measured at the output of the interferometer in \fref{fig:2 mode unitary transformation polarization correlation example}(b).
We emphasize that no entanglement process occurs in the interferometer; therefore the proposed scheme is not an entangling gate. Nonetheless, the measurement of the correlation \eref{eq:PcNOT2} between the photon-number fluctuations at the two output ports allow us to simulate a CNOT-gate operation. As a ``bonus'', we find that the correlation signal can be enhanced on demand by increasing the square $\overline{r}^2$ of the source mean-photon rate, making it robust against technological losses. 



As an example, if we fix the polarization angles $\phi_{\Ccurly}=\pi/4$ and $\phi_{\Tcurly}=0$ in the setup in \fref{fig:2 mode unitary transformation polarization correlation example}(b), the expectation value in \eref{eq:PcNOT2} reads
\begin{equation}
 \avg{\Delta n_{{\theta}_{C}} \Delta n_{{\theta}_{T}}} \propto \overline{r}^2 \cos^{2}\left(\theta_{C}-\theta_{T}\right),
 \nonumber 
 \label{eq:cpnf bell state example}
\end{equation}
simulating the $100 \%$-visibility correlations typical of a Bell state $\ket{\Phi^{+}}=\left(\ket{HH}+\ket{VV} \right)/\sqrt{2}$ even if no Bell state is produced. Indeed, neither an entangled source is used, as for example in the experiment of Sanaka, Kawahara and Kuga \cite{sanaka}, nor an entanglement process occurs in the interferometer as in a genuine CNOT gate operation. Nonetheless, an observer $T$ can provide to a separate observer $C$ the measured fluctuations $\Delta n_{{\theta}_{C}}$ in the number of photons for a polarization angle $\theta_{C}$ (in a given computational base) unknown to $T$, who can infer the value  of $\theta_{C}$ only based on his/her corresponding measurements of $\Delta n_{{\theta}_{T}}$ (with $\theta_{T}$ in the same computational base). Interestingly, differently from a genuine Bell state, the correlation between the measurements of the two observers $C$ and $T$ emerge \textit{only} from the expectation value of the product of the corresponding photon number fluctuations $\Delta n_{\theta_{C}}$ and $\Delta n_{\theta_{T}}$.

\begin{figure}
\centering
\includegraphics[scale=1]{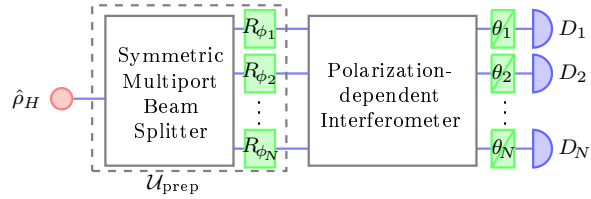}
\caption{$N$-order interference network with a thermal source based on: 1) Preparation of the $N$-channel initial state for arbitrary polarization angles $\phi_{i}$ ($i=1,\dots ,N$) with a generalized transformation $U_{prep}$ implemented by a symmetric $2 N$-port beam-splitter and $N$ half-wave plates; 2) $N$-channel polarization-dependent evolution consisting of polarization rotations and $\Pmc$-type transformations as in \fref{fig:2 mode unitary transformation polarization correlation example}(b); 3) Polarization-dependent correlation measurements in the photon-number fluctuations at the output of the network.}
\label{fig:3}
\end{figure}
\section{N-order Interference Networks} \label{sec:mpn}
One can finally generalize the second-order interference effect described in this paper to arbitrary orders $N$. 
For example, we consider an $N$-order interferometer as in \fref{fig:3} which generalizes the scheme in \fref{fig:2 mode unitary transformation polarization correlation example}(b). 
For this purpose, we first prepare the $N$-channel initial polarization state:   $H$-polarized thermal light impinges on a symmetric $2N$-port beam-splitter (generalization of the balanced beam splitter BS1 in \fref{fig:2 mode unitary transformation polarization correlation example}(b));  at each output port the polarization is then rotated, respectively, by arbitrary polarization angles  $\phi_{i}$, with $i=1,2,...,N$, by using $N$ half-wave plates. This prepared initial state propagates through a $2N$-port polarization-dependent interferometer, 
consisting of polarization rotations and  $\Pmc$-type transformations as in \fref{fig:2 mode unitary transformation polarization correlation example} (b). 
Finally, the polarization-dependent correlation between the photon-number fluctuations is measured at the $N$ output ports at approximately equal detection times. 

We emphasize that for $N=2$ one may also measure this correlation \eref{eq:PcNOT2} by subtracting the product of the independent light intensities at the two detectors from the standard second-order correlation function. Differently, in higher order interferometric networks it may be useful to directly measure the correlation between the photon-number fluctuations.

Experimentally, the precision in measuring the expectation value of the product of the photon-number fluctuations $\Delta n_{\theta_d}$ at each output port $d=1,2,...,N$ of a given network  depends on the number $n$ of performed measurements. For large numbers $n$ of measurements the distribution of the measured mean values is normal around the expectation value with an indetermination given by the indetermination in the  distribution of the measurement outcomes normalized by the root of $n$.   Evidently, the indetermination in the  distribution of the measurement outcomes depends intrinsically on the given $N$-order interference network. In general, a scaling in the number of resources typical of a quantum network with genuine entanglement cannot be achieved by the scheme described here. Nonetheless, an $N$-ordered interferometer as in \fref{fig:3} could be used to simulate experimentally $100\%$-visibility correlations typical of entangled states of $N$ qubits \cite{Peng2015}, such as GHZ states, as well as small-scale quantum circuits and algorithms. 




\section{Discussion} \label{sec:disc}
We demonstrated a novel interference phenomenon emerging from the fundamental nature of multipath correlations with a thermal source. 
In particular, we introduced a novel interferometer (see \fref{fig:2 mode unitary transformation polarization correlation example}(a)) where full correlations between the interferometric paths $S_C$ and $S_T$ ($L_C$ and $L_T$ )  emerge at the interferometer output from measuring the correlation between the fluctuations in the number of detected photons at the two output ports.
We showed how the interference between the two pairs $(S_C,S_T)$ and $(L_C,L_T)$ of correlated optical paths occurs even if the time delay imprinted by each path in one pair with respect to each path in the other pair is much larger than the source coherence time. We also pointed out the differences and the similarities between this interferometer and the well known Franson interferometer where an entangled two-photon source is used instead of a thermal source.

The interference effect demonstrated here can be easily observed experimentally: 1) It relies on one of the most natural sources which can be easily simulated in a laboratory by using laser light impinging on a fast rotating ground glass \cite{Arecchi196627}; 2) The calibration of the interferometric paths and of the detection times (see \eref{eq:Cnot conditions 02aa}, \eref{eq:Cnot conditions 02a} and \eref{eq:Cnot conditions 01c}) can be easily achieved in the case of a thermal source, where the coherence time can range from the order of $ns$ to $\mu s$. 
In an analogous way, multipath correlation interference can be also experimentally observed in the spatial domain with a spatial-mode dependent thermal source \cite{Cassano2016}.

In conclusion, this interesting phenomenon provides a deeper fundamental understanding of the physics of coherence, multipath correlations and interference using a thermal source.
Furthermore, it can be used to implement correlations in different degrees of freedom without recurring to entanglement processes.
As an example, we demonstrated the polarization-encoded simulation of a CNOT logic operation.
We also showed how this second-order interference effect can be extended to arbitrary orders $N$, leading to the interference of more general configurations of correlated paths. This could be used to simulate on-demand $100\%$-visibility correlations typical of entangled states of $N$ qubits, with possible applications in high precision metrology and imaging \cite{Valencia, Oppel_2012, crespi2012, Pearce2015}, and in the development of novel optical algorithms \cite{tamma_analogue_2015-1, tamma_analogue_2015, PRARapidFact, JOMFact, Tamma_2010}. 
\appendix

\ack
V.T. would like to thank M. Cassano, M. D'Angelo, P. Facchi, J. Fan, J. Franson, M. Freyberger, A. Garuccio, Y.-H. Kim, S. Laibacher, A. Migdall,  S. Pascazio, T. Peng, T. Pittman, W.P. Schleich, and Y.H. Shih for fruitful conversations about the proposed scheme.  V.T. is also grateful to M. Cassano, T. Peng and Y.H. Shih for discussions, during his visit at UMBC in 2014, about the experimental realization of this proposal in the spatial domain in Y.H. Shih's laboratory (paper in preparation). 

V.T. acknowledges the support of the German Space Agency DLR with funds provided by the Federal Ministry of Economics and Technology (BMWi) under grant no. DLR 50 WM 1136.

\appendix

\section{Correlation between the photon-number fluctuations for the interferometer in \fref{fig:2 mode unitary transformation polarization correlation example}(a) }\label{sec:appA}
We find here the explicit expression of the expectation value (in \eref{eq:corrfluc})
\begin{equation}
\avg{\Delta n_{C} \Delta n_{T}}\propto\abssq{G^{(1)}(t_{C},t_{T})}
\label{eq:appA_1}
\end{equation}
of the product of the photon-number fluctuations by calculating the first-order correlation function (in \eqref{eq:first-order-correlation})
\begin{equation}
G^{(1)}(t_{C},t_{T})\defeq\tr \left[\hat{\rho}_{H} {\hat{E}}_{C}^{(-)}\!(t_{C})  {\hat{E}}_{T}^{(+)}\!(t_{T})  \right],
\label{eq:app1_g1}
\end{equation}
with the electric field operators
\begin{equation}
{\hat{E}}_{C}^{(-)}(t_{C})=-\frac{K}{\sqrt{2}} \!\!\int\!\! \difd \omega\; \e^{i\omega t_{C}}\left[\e^{-\ii \omega L_{C}/c} + \e^{-\ii \omega S_{C}/c} \right] \;
\hat{a}_{A}^{(H)\dagger}(\omega),
\label{eq:app1_efield}
\end{equation}
and
\begin{equation}
{\hat{E}}_{T}^{(+)}(t_{T})=\frac{\ii K}{\sqrt{2}} \!\!\int\!\! \difd \omega\; \e^{-i\omega t_{T}}\left[\e^{\ii \omega L_{T}/c} + \e^{\ii \omega S_{T}/c} \right] \;
\hat{a}_{A}^{(H)}(\omega),
\label{eq:app1_efield2}
\end{equation}
where $K$ is a constant. By inserting \eref{eq:app1_efield} and \eref{eq:app1_efield2} into \eref{eq:app1_g1}, and using the definition (in \eref{eq:def_eff_det_times}) of the effective detection times
\begin{equation}
t_d^{(l_d)} = t_{d} - \frac{l_{d}}{c} \notag,
\end{equation}
we obtain  
\begin{align}
G^{(1)}(t_{C},t_{T})
= -\frac{\ii K^2}{2}&\left[  \tr \left(\hat{\rho}_{H}\!\! \int \!\! \difd \omega \!\! \int \!\! \difd \omega^{\prime} \e^{\ii(\omega t_{C}^{(S_{C})}-\omega^{\prime} t_{T}^{(S_{T})})}\hat{a}^{(H)\dagger}_{A}(\omega)\hat{a}_{A}^{(H)}(\omega^{\prime}) \right) \right. \notag \\ 
&\left.  +\tr \left(\hat{\rho}_{H}\!\! \int \!\! \difd \omega \!\! \int \!\! \difd \omega^{\prime} \e^{\ii\omega  t_{C}^{(S_{C})}-\omega^{\prime} t_{T}^{(L_{T})})}\hat{a}^{(H)\dagger}_{A}(\omega)\hat{a}_{A}^{(H)}(\omega^{\prime}) \right) \right. \notag \\ 
&\left.  +\tr \left(\hat{\rho}_{H}\!\! \int \!\! \difd \omega \!\! \int \!\! \difd \omega^{\prime} \e^{\ii(\omega t_{C}^{(L_{C})}-\omega^{\prime} t_{T}^{(S_{T})})}\hat{a}^{(H)\dagger}_{A}(\omega)\hat{a}_{A}^{(H)}(\omega^{\prime}) \right)\right. \notag \\ 
&\left.  +\tr \left(\hat{\rho}_{H}\!\! \int \!\! \difd \omega \!\! \int \!\! \difd \omega^{\prime} \e^{\ii(\omega t_{C}^{(L_{C})}-\omega^{\prime} t_{T}^{(L_{T})})}\hat{a}^{(H)\dagger}_{A}(\omega)\hat{a}_{A}^{(H)}(\omega^{\prime}) \right) \right] .
\nonumber 
\end{align}
By using the
property \cite{tammalaibacherthermal2014,Glauber2007}
\begin{align}
 \tr \left(\hat{\rho}_{H} \!\! \int \!\! \difd \omega \!\! \int \!\! \difd \omega^{\prime} \e^{\ii(\omega t_{1}-\omega^{\prime} t_{2})}\hat{a}^{(H)\dagger}_{A}(\omega)\hat{a}_{A}^{(H)}(\omega^{\prime}) \right) =\overline{r}\e^{\ii \omega_{0}(t_{1}-t_{2})}\e^{-(t_{1}-t_{2})^{2}\Delta\omega^{2}/2}
\nonumber 
\end{align}
for the multimode thermal state $\hat{\rho}_{H}$ in \eref{eq:Density_op_Thermal_B_Two_Mode}, we obtain
\begin{align}
G^{(1)}(t_{C},t_{T})= -\frac{\ii K^2}{2}\overline{r} &\left[ \e^{\ii \omega_{0}[t_{C}^{(S_{C})}-t_{T}^{(S_{T})}]}\e^{-[t_C^{(S_C)}- t_T^{(S_T)}]^2 \Delta \omega^2 /2} +\e^{\ii \omega_{0}[t_{C}^{(S_{C})}-t_{T}^{(L_{T})}]}\e^{-[t_C^{(S_C)}- t_T^{(L_T)}]^2 \Delta \omega^2 /2} \right. \notag \\ 
&\left. +\e^{\ii \omega_{0}[t_{C}^{(L_{C})}-t_{T}^{(S_{T})}]}\e^{-[t_C^{(L_C)}- t_T^{(S_T)}]^2 \Delta \omega^2 /2}  +\e^{\ii \omega_{0}[t_{C}^{(L_{C})}-t_{T}^{(L_{T})}]}\e^{-[t_C^{(L_C)}- t_T^{(L_T)}]^2 \Delta \omega^2 /2}  \right],
\notag
\end{align}
which can be rewritten as
\begin{equation}
G^{(1)}(t_{C},t_{T}) = \sum_{l_{C},l_{T}} G^{(l_{C},l_{T})}(t_{C},t_{T}),
\label{eq:appA_star}
\end{equation}
with the contributions
\begin{equation}
G^{(l_{C},l_{T})}(t_{C},t_{T}) = -\frac{\ii K^2}{2}\overline{r}\e^{\ii \omega_{0}[t_{C}^{(l_{C})}-t_{T}^{(l_{T})}]}\e^{-[t_C^{(l_C)}- t_T^{(l_T)}]^2 \Delta \omega^2 /2}
\notag
\end{equation}
for all possible pairs $(l_{C},l_{T})=(L_{C},L_{T}),(S_{C},S_{T}),(L_{C},S_{T}),(S_{C},L_{T})$ of optical paths.
By substituting \eqref{eq:appA_star} in \eqref{eq:appA_1} we finally find the expression of $\avg{\Delta n_{C} \Delta n_{T}}$ in \eqref{eq:corrfluct}. 

\section{Polarization-dependent correlation between the photon-number fluctuations for the interferometer in \fref{fig:2 mode unitary transformation polarization correlation example}(b)}\label{sec:appB}
We calculate here the expectation value (in \eref{eq:corrfluc2}) 
\begin{equation}
\avg{\Delta n_{{\theta}_{C}} \Delta n_{{\theta}_{T}}}_{C,T}\propto\abssq{G^{(1)}_{{\theta}_{C},{\theta}_{T}}(t_{C},t_{T})}
\label{eq:A1}
\end{equation}
for the product of the  photon-number fluctuations, with the polarization-dependent first-order correlation function  
\begin{equation}
G^{(1)}_{{\theta_{C}},{\theta_{T}}}(t_{C},t_{T})\defeq\tr \left[\hat{\rho}_{H} \left( \boldsymbol\theta_{C} \cdot \boldsymbol{\hat{E}}_{C}^{(-)}\!(t_{C})  \right)\left( \boldsymbol\theta_{T} \cdot \boldsymbol{\hat{E}}_{T}^{(+)}\!(t_{T}) \right) \right].
\label{eq:A2}
\end{equation}
The electric field operator $\boldsymbol{\hat{E}}^{(+)}_{d}\!(t_{d})$, with $d=C,T$, for direction of propagation perpendicular to the H-V plane is given in the narrow bandwidth approximation by the operator
\begin{equation}
\boldsymbol{\hat{E}}_{d}^{(+)}(t_{d})=\ii K \!\!\int\!\! \difd \omega\; \e^{-i\omega t_{d}}\Mmc_{d,A}(\omega) \;
\begin{pmatrix}
\hat{a}_{A}^{(H)}(\omega) \\ \hat{a}_{A}^{(V)}(\omega)
\end{pmatrix},
\label{eq:posfoout}
\end{equation}
with a constant $K$, the elements $\Mmc_{d,A}=\Mmc_{C,A},\Mmc_{T,A}$ in the first column of the total interferometer matrix $\Mmc (\omega)$ in \eref{eq:4by4transformation}, and the annihilation operators $\hat{a}_{A}^{(H)}(\omega)$ and $\hat{a}_{A}^{(V)}(\omega)$ at the only  port $A$ where a source is placed, while $\boldsymbol{\hat{E}}^{(-)}_{d}\!(t_{d})$ is its respective Hermitian conjugate. 


By defining
\begin{equation}
A_{C}\defeq
\left( \cos\theta_{C}\quad\sin\theta_{C}\right)
\,{\Mmc}_{C,A}(\omega)\,
\begin{pmatrix}
1 \\ 0
\end{pmatrix}
=\frac{\ii}{\sqrt{2}}\left(\cos\theta_{C}\cos\phi_{\Ccurly}\e^{\ii\omega S_{C}/c}+\sin\theta_{C}\sin\phi_{\Ccurly}\e^{\ii\omega L_{C}/c}\right),
\notag
\end{equation}
and
\begin{align}
A_{T}\defeq &
\left( \cos\theta_{T}\quad\sin\theta_{T}\right)
\,{\Mmc}_{T,A}(\omega)\, 
\begin{pmatrix}
1 \\ 0
\end{pmatrix}
\notag \\
=&\frac{1}{2\sqrt{2}}\left[\cos\theta_{T}\left(\cos\phi_{\Tcurly}\e^{\ii\omega S_{T}/c}+\sin\phi_{\Tcurly}\e^{\ii\omega L_{T}/c} \right) + \sin\theta_{T}\left(\sin\phi_{\Tcurly}\e^{\ii\omega S_{T}/c} + \cos\phi_{\Tcurly}\e^{\ii\omega L_{T}/c} \right) \right],
\notag
\end{align}
it is useful to introduce the \textit{effective} field operators
\begin{equation}
{\hat{\mathcal{E}}}_{d}^{(+)}\!(t_{d})
\defeq
\ii K\!\!\int\!\! \difd \omega \; A_{d}\; \e^{-i\omega t_{d}}{\hat{a}}_{A}^{(H)}(\omega),
\label{eq:efieldeff} 
\end{equation}
with $d=C,T$. Indeed, given the fixed polarization $H$ of the thermal light produced by the source in the port $A$, equation \eref{eq:A2} can be rewritten by using \eref{eq:posfoout} and \eref{eq:efieldeff} as
\begin{equation}
G^{(1)}_{{\theta_{C}},{\theta_{T}}}(t_{C},t_{T})\defeq\tr \left[\hat{\rho}_{H} \left( \mathcal{\hat{E}}_{C}^{(-)}\!(t_{C})  \right)\left( \mathcal{\hat{E}}_{T}^{(+)}\!(t_{T}) \right) \right].
\notag
\end{equation}
 By direct substitution and using again the definition (in \eref{eq:def_eff_det_times}) of the effective detection times
\begin{equation}
t_d^{(l_d)} \defeq t_{d} - \frac{l_{d}}{c},
\end{equation} 
with $l_{d}=L_{d},S_{d}$, we obtain
\begin{align}
G^{(1)}_{{\theta_{C}},{\theta_{T}}}(t_{C},t_{T}) 
= -\frac{\ii K^2}{4}&\left[ \vphantom{\int} \cos\theta_{C}\cos\phi_{\Ccurly}\left(\cos\theta_{T}\cos\phi_{\Tcurly} + \sin\theta_{T}\sin\phi_{\Tcurly}\right)\right. \notag \\
&\left. \quad \times\tr \left(\hat{\rho}_{H}\!\! \int \!\! \difd \omega \!\! \int \!\! \difd \omega^{\prime} \e^{\ii(\omega t_{C}^{(S_{C})}-\omega^{\prime} t_{T}^{(S_{T})})}\hat{a}^{(H)\dagger}_{A}(\omega)\hat{a}_{A}^{(H)}(\omega^{\prime}) \right)\right. \notag \\
&\left.+\cos\theta_{C}\cos\phi_{\Ccurly}\left(\cos\theta_{T}\sin\phi_{\Tcurly} +\sin\theta_{T}\cos\phi_{\Tcurly}\right)\right. \notag \\
&\left. \quad \times\tr \left(\hat{\rho}_{H}\!\! \int \!\! \difd \omega \!\! \int \!\! \difd \omega^{\prime} \e^{\ii(\omega t_{C}^{(S_{C})}-\omega^{\prime} t_{T}^{(L_{T})})}\hat{a}^{(H)\dagger}_{A}(\omega)\hat{a}_{A}^{(H)}(\omega^{\prime}) \right) \right. \notag \\ 
&\left.+\sin\theta_{C}\sin\phi_{\Ccurly}\left(\cos\theta_{T}\cos\phi_{\Tcurly} +\sin\theta_{T}\sin\phi_{\Tcurly}\right)\right. \notag \\
&\left. \quad \times\tr \left(\hat{\rho}_{H}\!\! \int \!\! \difd \omega \!\! \int \!\! \difd \omega^{\prime} \e^{\ii(\omega t_{C}^{(L_{C})}-\omega^{\prime} t_{T}^{(S_{T})})}\hat{a}^{(H)\dagger}_{A}(\omega)\hat{a}_{A}^{(H)}(\omega^{\prime}) \right) \right. \notag \\
&\left.+\sin\theta_{C}\sin\phi_{\Ccurly}\left(\cos\theta_{T}\sin\phi_{\Tcurly} +\sin\theta_{T}\cos\phi_{\Tcurly}\right)\right. \notag \\
&\left. \quad \times\tr \left(\hat{\rho}_{H}\!\! \int \!\! \difd \omega \!\! \int \!\! \difd \omega^{\prime} \e^{\ii(\omega t_{C}^{(L_{C})}-\omega^{\prime} t_{T}^{(L_{T})})}\hat{a}^{(H)\dagger}_{A}(\omega)\hat{a}_{A}^{(H)}(\omega^{\prime}) \right)  \right].
\nonumber 
\end{align}
By using again the
property \cite{tammalaibacherthermal2014,Glauber2007}
\begin{align}
 \tr \left(\hat{\rho}_{H} \!\! \int \!\! \difd \omega \!\! \int \!\! \difd \omega^{\prime} \e^{\ii(\omega t_{1}-\omega^{\prime} t_{2})}\hat{a}^{(H)\dagger}_{A}(\omega)\hat{a}_{A}^{(H)}(\omega^{\prime}) \right) =\overline{r}\e^{\ii \omega_{0}(t_{1}-t_{2})}\e^{-(t_{1}-t_{2})^{2}\Delta\omega^{2}/2}
\nonumber 
\end{align}
for the multimode thermal state $\hat{\rho}_{H}$ in \eref{eq:Density_op_Thermal_B_Two_Mode},
we obtain
\begin{align}
G^{(1)}_{{\theta_{C}},{\theta_{T}}}(t_{C},t_{T})=-\frac{\ii K^{2}}{4}\overline{r}&\left[ \cos\theta_{C}\cos\phi_{\Ccurly}\cos\left(\theta_{T}-\phi_{\Tcurly}\right) \e^{\ii \omega_{0}[t_{C}^{(S_{C})}-t_{T}^{(S_{T})}]}\e^{-[t_{C}^{(S_{C})}-t_{T}^{(S_{T})}]^{2}\Delta\omega^{2}/2}\right. \notag \\
&\left.+\cos\theta_{C}\cos\phi_{\Ccurly}\sin\left(\theta_{T}+\phi_{\Tcurly}\right)\e^{\ii \omega_{0}[t_{C}^{(S_{C})}-t_{T}^{(L_{T})}]}\e^{-[t_{C}^{(S_{C})}-t_{T}^{(L_{T})}]^{2}\Delta\omega^{2}/2} \right. \notag \\ 
&\left.+\sin\theta_{C}\sin\phi_{\Ccurly}\cos\left(\theta_{T}-\phi_{\Tcurly}\right)\e^{\ii \omega_{0}[t_{C}^{(L_{C})}-t_{T}^{(S_{T})}]}\e^{-[t_{C}^{(L_{C})}-t_{T}^{(S_{T})}]^{2}\Delta\omega^{2}/2} \right. \notag \\
&\left.+\sin\theta_{C}\sin\phi_{\Ccurly}\sin\left(\theta_{T}+\phi_{\Tcurly}\right)\e^{\ii \omega_{0}[t_{C}^{(L_{C})}-t_{T}^{(L_{T})}]}\e^{-[t_{C}^{(L_{C})}-t_{T}^{(L_{T})}]^{2}\Delta\omega^{2}/2}  \right].
\label{eq:first-order-correlation-app 2} 
\end{align}
Finally, by applying the conditions (in \eref{eq:8a} and \eref{eq:8})
\begin{align}
 \abs{t_{C}^{(L_{C})}-t_{T}^{(S_{T})}}{\Delta\omega}\gg 1, \quad
 \abs{t_{T}^{(L_{T})}-t_{C}^{(S_{C})}}{\Delta\omega}\gg 1 ,
\nonumber \\
\abs{t_{C}^{(L_{C})}-t_{T}^{(L_{T})}}{\Delta\omega}\ll 1,  \quad
\abs{t_{C}^{(S_{C})}-t_{T}^{(S_{T})}}{\Delta\omega}\ll 1 ,
\nonumber 
\end{align}
 \eref{eq:first-order-correlation-app 2} reduces to 
\begin{equation*}
 G_{\theta_{C}, \theta_{T}}^{(1)}(t_{C},t_{T})=G_{\theta_{C}, \theta_{T}}^{(L_{C},L_{T})}(t_{C},t_{T})+G_{\theta_{C}, \theta_{T}}^{(S_{C},S_{T})}(t_{C},t_{T}),
\end{equation*}
with the two contributions
\begin{align}
G_{\theta_{C}, \theta_{T}}^{(S_{C},S_{T})}(t_{C},t_{T})&= -\ii\frac{K^{2}}{4}\overline{r}\cos\phi_{\Ccurly}\cos\theta_{C} \cos\left(\phi_{\Tcurly}-\theta_{T}\right) \e^{\ii \omega_{0} \left[(t_{C}^{(S_{C})} - t_{T}^{(S_{T})} \right]},\nonumber 
\\
G_{\theta_{C}, \theta_{T}}^{(L_{C},L_{T})}(t_{C},t_{T})&= -\ii\frac{K^{2}}{4}\overline{r}\sin\phi_{\Ccurly}\sin\theta_{C} \sin\left(\phi_{\Tcurly}+\theta_{T}\right) \e^{\ii \omega_{0} \left[(t_{C}^{(L_{C})}-t_{T}^{(L_{T})}\right]}. \nonumber 
\end{align} 
Thereby, \eref{eq:A1} reads
\begin{equation}
\avg{\Delta n_{{\theta}_{C}} \Delta n_{{\theta}_{T}}}_{C,T}\propto\abssq{G_{\theta_{C} \theta_{T}}^{(L_{C},L_{T})}(t_{C},t_{T})+G_{\theta_{C} \theta_{T}}^{(S_{C},S_{T})}(t_{C},t_{T})},
\notag
\end{equation}
which, by introducing the relative phase (in (\ref{eq:limit of omega 0}))
\begin{equation}
\varphi_{L-S}\defeq \frac{\omega_{0}}{c}\left[(L_{C}-L_{T})-(S_{C}-S_{T} )\right]
\nonumber 
\label{eq:limit of omega 0 2},
\end{equation} 
finally reduces to
\begin{align}
\avg{\Delta n_{{\theta}_{C}} \Delta n_{{\theta}_{T}}}_{C,T}\propto \overline{r}^2 \left|\cos\phi_{\Ccurly}\cos\theta_{C} \cos\left(\phi_{\Tcurly}-\theta_{T}\right) + \e^{\ii \varphi_{L-S}} \sin\phi_{\Ccurly}\sin\theta_{C} \sin\left(\phi_{\Tcurly}+\theta_{T}\right)  \right|^{2},
\notag
\end{align}
as in \eref{eq:pnf coh gen 2}.

\section*{References}

%

\providecommand{\newblock}{}


\end{document}